\documentclass{article}

\usepackage{arxiv}

\usepackage[utf8]{inputenc} % allow utf-8 input
\usepackage[T1]{fontenc}    % use 8-bit T1 fonts
\usepackage{hyperref}       % hyperlinks
\usepackage{url}            % simple URL typesetting
\usepackage{booktabs}       % professional-quality tables
\usepackage{amsfonts}       % blackboard math symbols
\usepackage{nicefrac}       % compact symbols for 1/2, etc.
\usepackage{microtype}      % microtypography
\usepackage{lipsum}
\usepackage{graphicx}

\title{Convolutional Neural Networks for Estimation of Myelin Maturation in Infant Brain}

\author{
  Akihiko Wada1, Yuya Saito1,2, Ryusuke Irie1,3, Koji Kamagata1, Tomoko Maekawa1,3, Shohei Fujita1\\
  Akifumi Hagiwara1,3, Kanako Kumamaru1, Michimasa Suzuki1\\
  Atsushi Nakanishi1, Masaaki Hori1,Toshiaki Shimizu3 and Shigeki Aoki1 \thanks{Department of Radiology, Juntendo University School of Medicine,   1-2-1, Hongo, Bunkyo-ku, Tokyo, Japan, 113-8421} \\
 1 Department of Radiology, Juntendo University School of Medicine \\
 2 Department of Radiological Sciences, Graduate School of Human Health Sciences,\\  Tokyo Metropolitan University  \\
 3 Department of Radiology, Graduate School of Medicine, The University of Tokyo  \\
 4 Department of Pediatrics, Juntendo University School of Medicine  \\
  \texttt{a-wada@juntendo.ac.jp} \\
}

\begin{document}
\maketitle

\begin{abstract}
\
 Myelination plays an important role in the neurological development of infant brain and MRI can visualize the myelination extension as T1 high and T2 low signal intensity at white matter. We tried to construct a convolutional neural network (CNN) machine-learning model to estimate the myelination. Eight layers CNN architecture was constructed to estimate the subject’s age with T1 and T2 weighted image at 5 levels associated with myelin maturation in 119 subjects up to 24 months. CNN model learned with all age dataset revealed a strong correlation between the estimated age and the corrected age and the coefficient of correlation, root mean square error and mean absolute error was 0.81, 3.40 and 2.28. Moreover, the adaptation of ensemble learning models with two datasets 0 to 16 months and 8 to 24 months improved that to 0.93, 2.12 and 1.34. Deep learning can be adaptable to myelination estimation in infant brain.
\end{abstract}

% keywords can be removed
% \keywords{First keyword \and Second keyword \and More}

\section{Introduction}
\
Myelination plays a major role in the neurological development of infants. 
In the central nervous system, oligodendrocytes, which are myelinogenic cells, wrap their cytoplasm around neuronal axons, forming a multilayered structure. \cite{Dobbing1973Quant} 
 The multiple myelin layers around an axon functions as an insulator, enabling extremely high-speed transmission of electric signals by saltatory conduction. 
This high-speed electrical signal transmission enables rapid neurological development of infants. Myelination is already initiated before birth; at birth, myelination is seen in the brain stem, cerebellar white matter, posterior limb of the internal capsule, and subcortical white matter of primary motor and sensory cortices. After birth, myelination progresses in the brain caudally to rostrally, posteriorly to anteriorly, and centrally to peripherally. Myelin maturation progresses on a monthly basis. (Figure 1)
Magnetic resonance imaging (MRI) visualizes myelin maturation in white matter as a high signal intensity on T1-weighted images and a low signal intensity on T2-weighted images. 	
\cite{Holland BA1986MRI}
 T1 / T2 shortening reflects an increase of lipid and protein and a decrease of water content associated with myelin maturation. \cite{McArdle CB1987Developmental} MRI evaluation of myelin maturation based on an alteration of MR signal is one of the objective evaluation methods of neurological development of infants. In evaluation of the neurological development of infants and children, T1-weighted images are useful in the early stage of myelination and T2-weighted images in the late stage. Clinically, radiologists and pediatricians often spend their precious time and effort to accurately predict neurological growth in children using brain MRI. 
\
A convolutional neural network (CNN) is a machine learning technique that excels in image recognition and classification. CNN architecture is composed of a multilayered convolutional layer and a classifier that can extract features of objects such as size, shape and density patterns from various levels of pixel units in each region. The superior CNN image recognition / classification ability may be able to contribute to the evaluation of myelination using brain MRI. The purpose of this study was to construct a machine learning model to estimate the age of infants from MR signal alteration of myelination using CNN architecture. 

\section{Materials and methods}
This retrospective non-invasive and non-intervention study was approved by the institutional review board of Juntendo University Hospital and opt-out alternatives was approved to obtaining informed consent for subjects.

\subsection{Participants}
The potentially eligible participants were 908 infants and children from birth to 24  months after birth, whom head MR imaging was performed in our institution from January 2014 to December 2015. 
Participants with obvious neurological symptoms or received some treatment intervention in the course of the subsequent two years were excluded. 133 participants were extracted under this exclusion condition. In addition, 14 participants with insufficient MR image quality with a motion artifact were excluded. As a result, 119 infants and children up to 2 years old, 40 boys and 79 girls were recruited for this study.(Figure 2) 
Each participants age was corrected by converting gestational age to a full 40 weeks, the corrected ages were from - 1.71 to 23.54 months.

\subsection{MR imaging}
Whole brain 2D T1- and T2-weighted images were obtained using a 3T MR unit (Achieva, Philips) in 95 subjects and a 1.5T MR unit (Magnetom Avanto, Siemens) in 24 subjects. The sequence parameters of MRI were as follows. Achieva, Philips 3T MR unit: T1WI; Turbo-Inversion recovery, TR/TE/TI: 2155/10/1000 msec, echo train length: 4, flip angle: 90 degrees; T2WI; turbo spin echo, TR/TE: 4000/80 msec, echo train length: 11, NEX: 2, FOV 180–230 mm, matrix 512x512, slice thickness: 5 mm. Magnetom Avanto, Siemens 1.5T MR unit: T1WI; spin echo; TR/TE: 530/11msec, echo train length: 1, flip angle: 80 degrees; T2WI; turbo spin echo; TR/TE: 4100/94, echo train length: 9, NEX: 2, FOV: 200–260 mm, matrix: 256x256, slice thickness: 5.5 mm.

\subsection{Data sets}
Five T1- and T2-weighted images at the levels of (a) the middle cerebellar peduncle of the pons, (b) midbrain, (c) internal capsule and splenium of corpus callosum, (d) central semiovale, and (e) subcortical white matter, were extracted by a neuroradiologist with 25 years of clinical experience (A.W) and adapted for input of machine learning. The corrected age from a subject's gestational age was adopted as the label of the supervised learning. We prepared two types of data sets to evaluate the effect of preprocessing of data sets on the accuracy of machine learning. One data set consisted of all 119 subjects and the other was composed of two components: from birth to 16 months and from 8 to 24 months. 

\subsection{Machine learning framework} 
Eight layers CNN architecture was consisting of 2 convolutional layers and 6 fully connected layers were developed using Neural network console ver. 1.2 (https://dl.sony.com/) on a Windows PC Intel Core i7 2.2 GH, with 32 GB memory and an NVIDIA GeForce GTX 1070 graphical processing unit. (Figure 2) The input image of 128x128 gray scale mediated by 2 convolution layers (5x5 convolutional kernel with valid padding, strides 1, 1 and out map was 16 and 8) and 2 Max Pooling layers (2x2 kernel and stride 2, 2) and 29 x 29 x8 feature maps was transmitted to Dense / fully connected layer (Dense) with 100 nodes. In each layers the rectified linear unit (ReLU) was applied as activation function. The output from 5 level T1 weighted images are concatenated and transmitted to 2 Dense layers with 100 nodes and concatenated with output from T2 weighted images and transmitted to Dense layers with 200, 50, 25 nodes with activation function ReLU and the learning model was trained to minimize the loss function Squared Error. Dense layer, which works as a classifier, was set in multiple levels at each MR image levels, T1 and T2weighted contrast levels and in all concatenated data. Two algorithms were constructed for infant age estimation using different two types of data sets. (Figure 3) One algorithm was a simple machine learning model that learned using the all-age data set. The other was an ensemble algorithm combining two learning models; one model learned with the data set from birth to 16 months and another model learned with the data set from 8 to 24 months. The latter split data set adoption is a configuration that takes into account the weight of myelin-related T1 / T2 signals up to 12 months and after that to 24 month. 
The following hyperparameters for training were common in both models: 1000 epochs; base learning rate for the untrained model, Adam (learning rate = 0.001, beta1 = 0.9, beta2 = 0.999, epsilon = 0.00000001). 

\subsection{Performance evaluation of the machine learning model} 
We compared the two machine learning models by Spearman’s rank-order correlation test, the root mean square error (RMSE) and the mean absolute error (MAE) to investigate correlations between the estimated age and a subject’s corrected age.

\section{Result}
There was a strong correlation between the estimated age by the two machine learning models and the corrected age of subjects. In model A, which learned with the all-age data set, the correlation coefficient, RMSE and MAE were 0.81 (p < 0.001), 3.40 and 2.28. In model B, which consisted of an ensemble of two data sets, learning showed a superior correlation coefficient, RMSE and MAE, of 0.93 (p < 0.001), 2.12 and 1.34, respectively. (Figure 4, Table 1)

\section{Discussion}
Myelination of axons enables high-speed transmission of electrical signals by fast saltatory conduction. Fast neural networks constructed with myelinated nerve fibers support rapid neurological growth and maturation of higher order cognitive functions in neonates and infants. In the central nervous system, the processes of oligodendrocytes, which are myelin-forming cells, become flattened and wind around axons like a roll of paper. The cytoplasm within the process gradually disappears and adjacent membranes adhere to each other, forming a sheath around axons. The myelin sheath is constructed of a multilayered spiral containing lipid layers and proteins. The lipid layers are composed of cholesterol, phospholipids and glycolipids and the ratio is approximately 4:3:2. 	
\cite{Barkovich AJ2000Concepts} Lipid layers support the saltatory conduction of myelinated fibers, allowing high-speed electrical transmission by acting as an insulator. Proteins maintain stability between the multiple membrane layers surrounding an axon. The protein component of myelin sheathes is composed of myelin basic protein and proteolipid protein, which are myelin-specific proteins. Associated with myelination of an infant's brain, the water content of white matter decreases and glycolipids and proteins of cell membranes constituting the myelin sheath increase.

By MRI, myelination of white matter leads to shortening of the T1 / T2 relaxation time; mature myelin exhibits a high T1 signal and low T2 signal compared to gray matter. 	
\cite{Holland BA1986MRI}
 The high T1 signal originates from an increase of cholesterol and galactocerebroside in the glycolipids that are the main components of the myelin sheath lipid layers. 	
\cite{Kucharczyk W1994Relaxivity, Koenig W1990Relaxometry}
 These molecules act on the magnetization transfer effect in adjacent free water to shorten the T1 relaxation time. Myelin maturation begins before birth; the full-term infant brain presents a high signal on the dorsal side of the brainstem, cerebellar peduncle and thalamus on T1-weighted images at birth. These MR signal alterations proceed through the brain from caudal to rostral, posteriorly to anteriorly and centrally to peripherally, along with myelin maturation. Three months after birth, the high T1 signal reaches the cerebellar white matter, the posterior limb of internal capsule, the periventricular white matter, and the subcortical white matter near the central sulcus. By 6 months, it reaches the optic radiation and the splenium of the corpus callosum and by 12 months, it reaches the entire white matter of the cerebral hemispheres. 

T1-weighted imaging is sensitive at an early stage of myelination, and its signal pattern visually exhibits an adult type approximately 8 months after birth. 	
\cite{Barkovich AJ1988Normal, Bird CR1989MR, Christophe C1900Mapping, van der Knaap1990MR, Martin E1988Developmental, Martin E1991MR}
 Conversely, alteration of the T2 signal is slower than the T1 signal and becomes significant in the late stage of myelin maturation. The T2 signal alteration is derived from the reduction of free water content that accompanies maturity of white matter beyond the biochemical changes. The water molecules inside myelin sheath and outside (at axon and interstitial tissue) directly affect the T2 signal intensity. 	
\cite{Barkovich AJ1988Normal, Bird CR1989MR, Christophe C1900Mapping, van der Knaap1990MR, Martin E1988Developmental, Martin E1991MR}
 As the density of the spiral structure of the myelin sheath increases, interstitial water molecules decrease and the hydrocarbon chains present in the lipid layers in the membrane increase the hydrocarbon bundle bands. The completion of alteration of the T2 signal of whole brain white matter is at around 2 years old. 	
\cite{Barkovich AJ2000Concepts}

CNN is a machine learning method suitable for image recognition that can capture the spatial features of an object. CNN was devised from experimental results of human visual cortex. 	
\cite{Fukushima K1982new,LeCun Y2015Deep}
 In the internal structure of CNN, the multilayered convolutional layers extract features of the object at various levels. In the convolutional layer, the matrix format input data is connected to the local receptive field and the convolutional process extracts the features of an object and creates a feature map. Multilayered convolutional layers extract the spatial features of subjects at different levels. The features of the object can be caught step by step from a simple element to a complex structure, from simple units such as pixels, lines and angles to structures such as a person’s eyes and nose, as well as more complex parts such as face and body. The feature map created by the convolutional layers is transformed into a vector and transmitted to the following fully connected layer. Multilayered fully connected layers perform a classification or regression analysis of the object as a classifier.

However, a proper pre-processing that makes it easy to recognize features of an object leads to improvement in the accuracy of machine learning. As effective preprocessing for image recognition and classification, selection of input data from a large amount of data, cutting of an object from an input image, and adjustment of contrast are worthy of mention. In this study, two pre-processing steps were applied to improve the accuracy of the learning model. As input data, 5 slice images that had characteristics of myelin maturation were selected from 20 or more whole-brain MR images. By this pre-processing, it was expected that the feature map created by CNN would be more related to myelination. In addition, considering that the mechanisms associated with MR signal changes during myelin maturation are different between T1 and T2 signals, two data sets with different target ages were created, and ensemble architecture of two machine learning models was constructed. Improvement of the accuracy of the learning model by these methods was shown by an improvement of the correlation coefficient and error loss.

As a next step, we are preparing for a pipeline that extracts the input data from whole brain MR images, estimates the myelination age by the ensemble machine learning model, and evaluates the delay of myelination in a subject's brain automatically.

There are some limitations in our study. The first limitation is that the number of samples was not large. Generally, in machine learning of image classification, sample sizes from several thousands to tens of thousands are applied. \cite{Krizhevsky A2012ImageNet}
 The number of samples in our study was less than this. However, it is equivalent to the median number of samples in a survey study of machine learning of medical images in brain regions within the last 5 years (mean 231, median 120). 	
\cite{Sakai K2018Machine}
 The second is that our estimation was performed with two static magnetic field strength MR units of 1.5T and 3T. 	
\cite{Bottomley PA1998review, Schmitz BL2005Advantages, Stanisz GJ2005T1}
 With 3T MRI, the T1 relaxation time is longer than with 1.5T and may affect myelination evaluation in the early stage of myelination. Creation of a data set and construction of a learning model corresponding to each static magnetic field intensity may improve the accuracy of myelination evaluation. 

\section{Conclusion}
The CNN machine learning model can estimate the age of infants and children with high accuracy from the MR signal intensity and is adaptable to the assessment of infant brain myelination. Our machine learning model will support the work of radiologists and pediatricians in evaluating the neurological development of infants.

\begin{figure}
  \centering
  \includegraphics{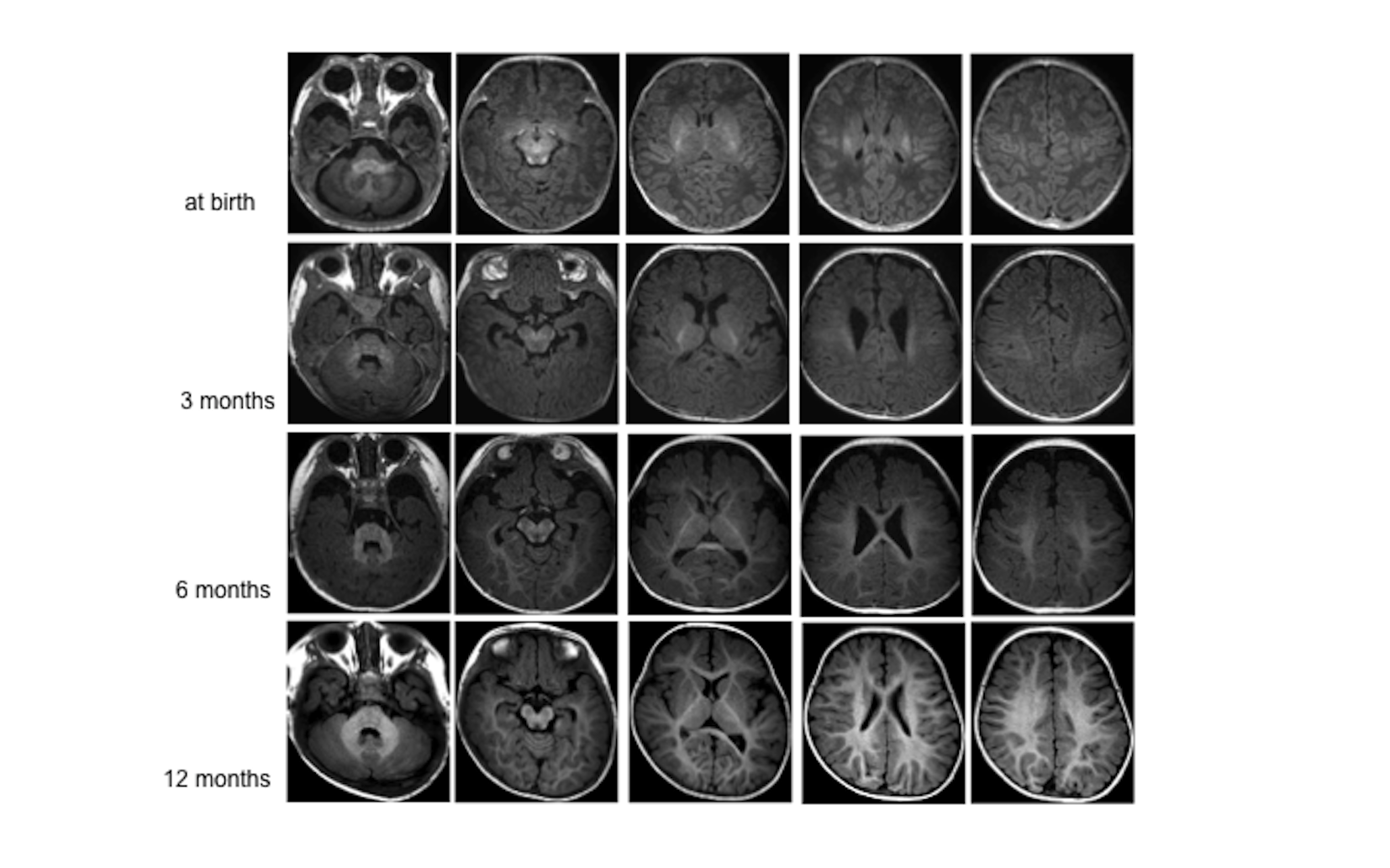}
  \label{fig.myelin}
    \caption{Normal myelin maturation on T1-weighted images}
      \flushleft
    The top row is full-term infant, the 2nd row is a child 3 months old, the 3rd row is 6 months old and the bottom row is 12 months old. The full-term infant brain presents a high signal on the dorsal side of the brain stem, cerebellar peduncle and thalamus on T1-weighted images at birth (top row). These MR signal alterations proceed through the brain from caudal to rostral, posteriorly to anteriorly and centrally to peripherally.
\end{figure}

\begin{figure}
  \centering
  \includegraphics{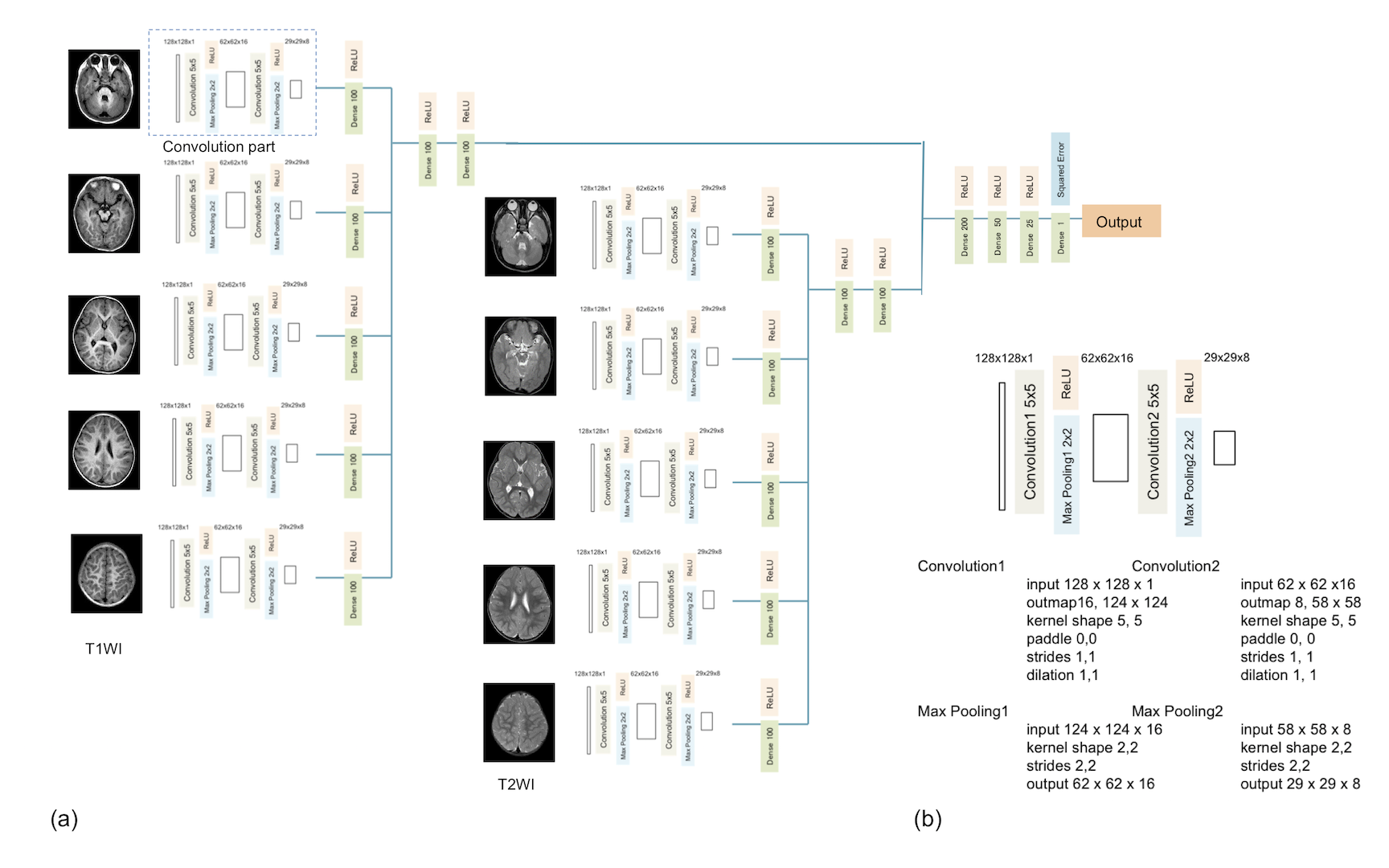}
  \label{fig.architecture}
    \caption{The architecture of convolution neural network.}
          \flushleft
    Eight layers convolution neural network with 5 levels T1 and T2 weighted images as input (a), and detail of the 2 layers convolutional part (b). In convolution part, 128x128 gray scale images mediated by 2 convolution layers; Conv1(5x5 convolutional kernel with valid padding, strides 1, 1 and out map was 16, Conv2 (5x5 convolutional kernel with valid padding, strides 1, 1 and out map was 8) and 2 Max Pooling layers (2x2 kernel and stride 2, 2) and 29 x 29 x8 feature maps was transmitted to following Dense / fully connected layer. In each layers the rectified linear unit (ReLU) was applied as activation function. The output from 5 convolution parts and following Dense / fully connected layer are concatenated and transmitted to 2 Dense layers with 100 nodes and concatenated with output from T2 weighted images and transmitted to Dense layers with 200, 50, 25 nodes with activation function ReLU and the learning model was trained to minimize the loss function Squared Error. 
\end{figure}

\begin{figure}
  \centering
  \includegraphics{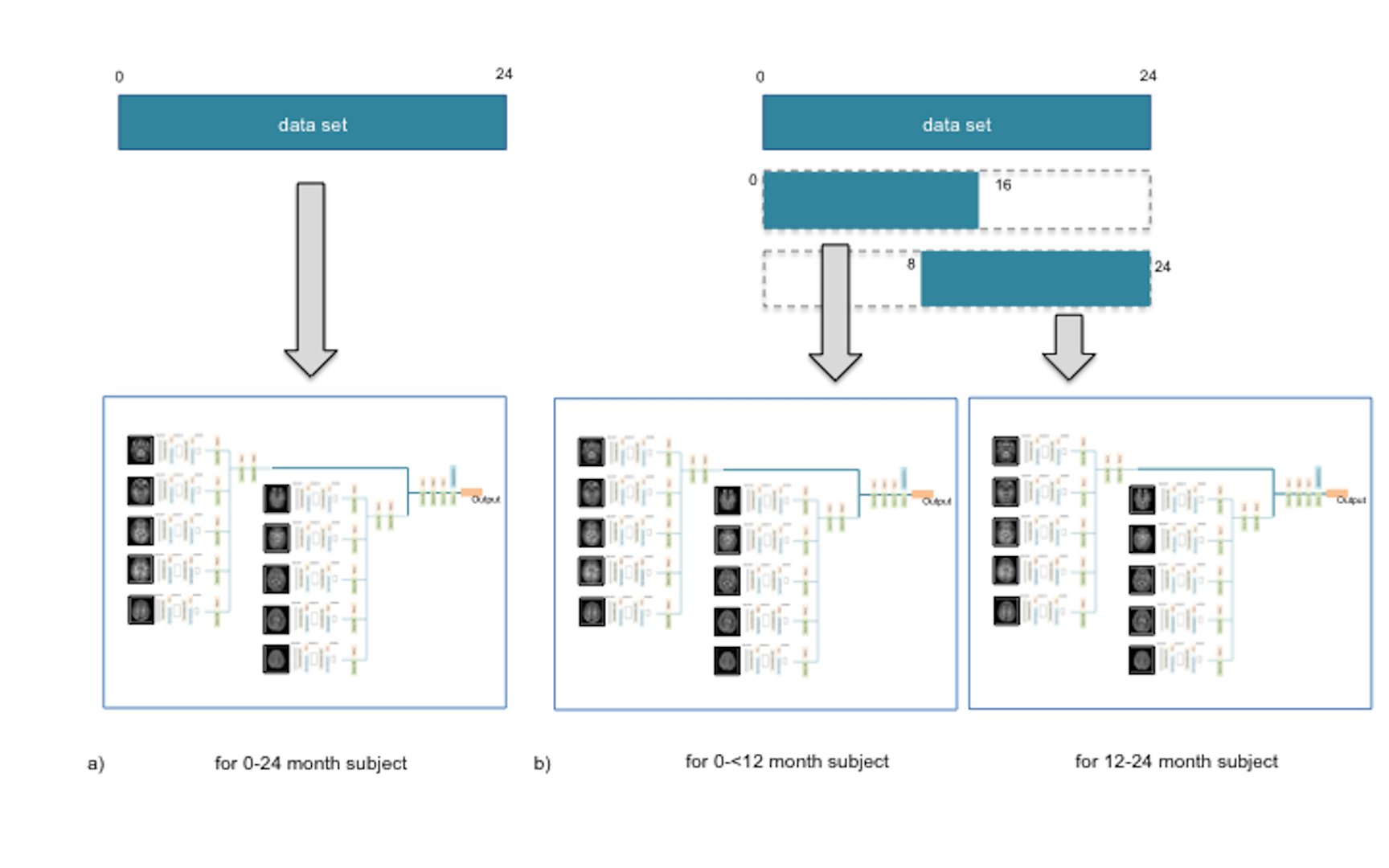}
  \label{fig.augumentation}
    \caption{Augmentation of dataset}
      \flushleft
    A simple learning model using all ages for a data set (a) and an ensemble learning model using two data sets from birth to 16 months and from 8 months to 24 months (b).
\end{figure}

\begin{figure}
  \centering
  \includegraphics{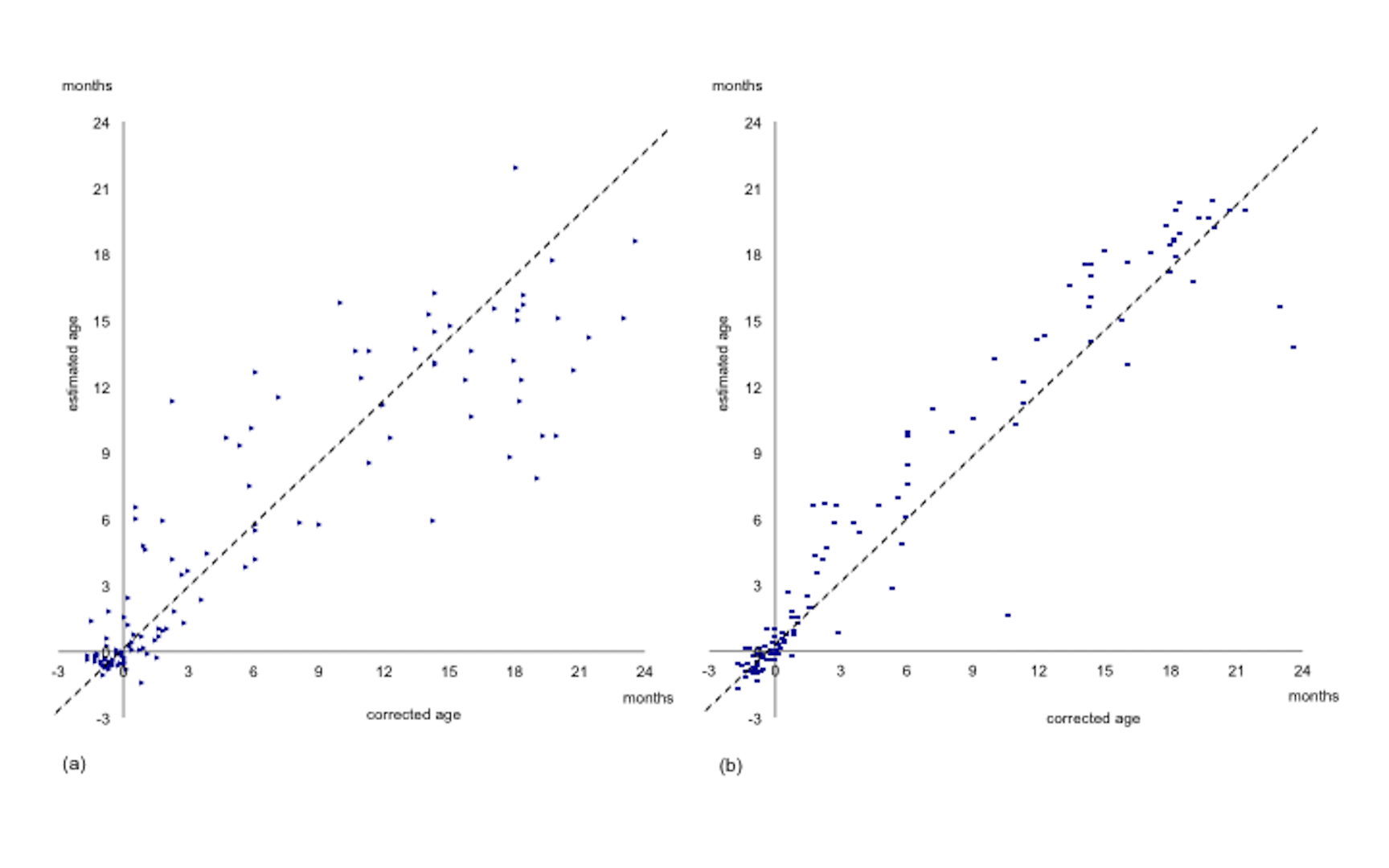}
  \label{fig.scatter}
    \caption{Scatter diagram of corrected age of subject and estimated age by machine learning in simple model}
      \flushleft
    Scatter diagram of corrected age of subject and estimated age by machine learning in simple model (a) and ensemble model (b).
\end{figure}

\begin{table}
 \caption{Estimation of two regression machine learning models}
  \label{tab:table}
  \centering
  \begin{tabular}{c c c} \hline
    \  & simple model & ensemble model \\    \hline \hline
    r & 0.901 & 0.964 \\
    RMSE & 3.40 & 2,.12 \\
    MAE & 2.26 & 1.34 \\ \hline

  \end{tabular}
 \centering
\end{table}

\bibliographystyle{unsrt}  


\begin{thebibliography}{1}


\bibitem{Dobbing1973Quant}
Dobbing J, Sands J 
\newblock Quantitative growth and development of human brain. Archives of Disease in Childhood, 48:757–767, 1973.

\bibitem{Holland BA1986MRI}
Holland BA, Haas DK, Norman D, Brant-Zawadzki M, Newton TH.
\newblock MRI of normal brain maturation. American Journal of Neuroradiology, 7:201–208, 1986.

\bibitem{McArdle CB1987Developmental}
McArdle CB, Richardson CJ, Nicholas DA, Mirfakhraee M, Hayden CK, Amparo EG. 
\newblock Developmental features of the neonatal brain: MR imaging. Part I. Gray-white matter differentiation and myelination. Radiology, 162:223–229,1987.

\bibitem{Barkovich AJ2000Concepts}
 Barkovich AJ.
\newblock Concepts of myelin and myelination in neuroradiology. Am J Neuroradiol. 21:1099–1109, 2000.

\bibitem{Kucharczyk W1994Relaxivity}
Kucharczyk W, Macdonald PM, Stanisz GJ, Henkelman RM.
\newblock Relaxivity and magnetization transfer of white matter lipids at MR imaging: importance of cerebrosides and pH. Radiology, 192:521–529,1994.

\bibitem{Koenig W1990Relaxometry}
Koenig SH, Brown RD, Spiller M, Lundbom N.
\newblock Relaxometry of brain: why white matter appears bright in MRI. Magn Reson Med. 14:482–495, 1990.

\bibitem{Barkovich AJ1988Normal}
Barkovich AJ, Kjos BO, Jackson DE Jr, Norman D.
\newblock Normal maturation of the neonatal and infant brain: MR imaging at 1.5 T. Radiology, 166:173–180, 1988.

\bibitem{Bird CR1989MR}
Bird CR, Hedberg M, Drayer BP, Keller PJ, Flom RA, Hodak JA.
\newblock MR assessment of myelination in infants and children: usefulness of marker sites. American Journal of Neuroradiology, 10:731–740, 1989. 

\bibitem{Christophe C1900Mapping}
Christophe C, Muller MF, Bal riaux D, et al. 
\newblock Mapping of normal brain maturation in infants on phase-sensitive inversion-recovery MR images. Neuroradiology, 2nd ed. Springer-Verlag. 32:173–178, 1990.

\bibitem{van der Knaap1990MR}
van der Knaap MS, VALK J.
\newblock MR imaging of the various stages of normal myelination during the first year of life. Neuroradiology, Springer-Verlag. 31:459–470, 1990.

\bibitem{Martin E1988Developmental}
Martin E, Kikinis R, Zuerrer M, et al.
\newblock Developmental Stages of Human Brain. J Comput Assist Tomogr, 12:917–922, 1988.

\bibitem{Martin E1991MR}
Martin E, Krassnitzer S, Kaelin P, Boesch C.
\newblock MR imaging of the brainstem: normal postnatal development. Neuroradiology. Springer-Verlag, 33:391–395, 1991.

\bibitem{Fukushima K1982new}
Fukushima K, Miyake S.
\newblock A new algorithm for pattern recognition tolerant of deformations and shifts in position. Pattern Recognition, 15:455–469, 1982. 

\bibitem{LeCun Y2015Deep}
LeCun Y, Bengio Y, Hinton G. 
\newblock Deep learning. Nature. Nature Publishing Group, 521:436–444, 2015.

\bibitem{Krizhevsky A2012ImageNet}
Krizhevsky A, Sutskever I, Hinton GE.
\newblock ImageNet Classification with Deep Convolutional Neural Networks. 1097–1105, 2012.

\bibitem{Sakai K2018Machine}
Sakai K, Yamada K.
\newblock Machine learning studies on major brain diseases: 5-year trends of 2014–2018. Jpn J Radiol. Springer Japan, 37:34–72, 2018.

\bibitem{Bottomley PA1998review}
Bottomley PA, Foster TH, Argersinger RE, Pfeifer LM.
\newblock A review of normal tissue hydrogen NMR relaxation times and relaxation mechanisms from 1-100 MHz: Dependence on tissue type, NMR frequency, temperature, species, excision, and age. Med Phys. John Wiley \& Sons, Ltd. 11:425–448. 1998. 

\bibitem{Schmitz BL2005Advantages}
Schmitz BL, Aschoff AJ, Hoffmann MHK, Grön G.
\newblock Advantages and pitfalls in 3T MR brain imaging: a pictorial review. American Journal of  Neuroradiology, 26:2229–2237, 2005.

\bibitem{Stanisz GJ2005T1}
Stanisz GJ, Odrobina EE, Pun J, et al.
\newblock T1, T2 relaxation and magnetization transfer in tissue at 3T. Magn Reson Med. John Wiley \& Sons, Ltd. 54:507–512, 2005.


\end{thebibliography}

\end{document}